# Realization of a transition between type-I and type-II Dirac semimetals in monolayers

Yuee Xie,[1,2] Yujiao Kang[1,2], Xiaohong Yan[1] and Yuanping Chen[1,2]*

[1]*School of Physics and Electronic Engineering, Jiangsu University, Zhenjiang, 212013, Jiangsu, China*
[2]*School of Physics and Optoelectronics, Xiangtan University, Xiangtan, Hunan, 411105, China*

## Abstract

The phase transition between type-I and type-II Dirac semimetals will reveal a series of significant physical properties because of their completely distinct electronic, optical and magnetic properties. However, no mechanism and materials have been proposed to realize the transition to date. Here, we propose that the transition can be realized in two-dimensional (2D) materials consisting of zigzag chains, by tuning external strains. The origination of the transition is that some orbital interactions in zigzag chains vary drastically with structural deformation, which changes dispersions of the corresponding bands. Two 2D nanosheets, monolayer PN and AsN, are searched out to confirm the mechanism by using first-principles calculations. They are intrinsic type-I or type-II Dirac materials, and transit to another type of Dirac materials by external strains. In addition, a possible routine is proposed to synthesize the new 2D structures.

*Correponding author: chenyp@ujs.edu.cn

Topological semimetals have a lot of classifications according to topological invariants, symmetries and other parameters. Depending on the dispersions of crossing bands, nodal-point semimetals (e.g. Dirac and Weyl semimetals) can be classified to two types[1-6]. In a type-I Dirac/Weyl semimetal, two degenerate/nondegenerate energy bands with *opposite-sign* slopes linearly cross at the Fermi level, and thus the Fermi surface consists of isolated type-I nodal points[7-13]. In a type-II Dirac/Weyl semimetal, the two crossing energy bands have *same-sign* slopes and form a strongly tilted cone. As a result, there exist electron and hole pockets on the Fermi level, and the type-II nodal point locates at boundaries of the pockets, which violates the Lorentz invariance required in high-energy physics[6, 14-19]. The two types of nodal points can transform from one to the other by a Lifshitz transition[20-29].

Although type-I and type-II nodal points share the same topological characteristics, they possess completely different electronic, optical and magnetic properties because of different tilting of the cones. For example, a type-I nodal-point semimetal exhibits negative magneto-resistance along all directions, however, the magneto-transport properties of type-II semimetals are expected to be extremely anisotropic and negative magneto-resistance is expected only along some certain directions[30, 31]; comparing with a type-I nodal-point semimetal, a type-II semimetal has a distinct response in its Landau level spectrum, i.e., the existence of a critical angle of the magnetic field, along which all Landau levels "collapse" into the same energy, giving rise to a large density of states[32]; in addition, the tilting of the cone has an important effect on the optical response and electronic transport, and thus leads to different Hall conductivity, valley polarization, Andreev reflection, Klein tunneling, and the anomalous Nernst effect[33, 34].

The significantly different properties between type-I and type-II Dirac/Weyl points indicate that a series of transitions will occur if a type-I nodal point is tuned to be a type-II one. Especially there exists a critical phase between type-I and type-II Dirac/Weyl points, named as critical nodal point, which is formed by crossing of a degenerate/nondegenerate flat band and a degenerate/nondegenerate dispersive band[21]. It has a tremendous density of state at the Fermi level, and thus may induce strong correlated effect and superconducting phenomena[35, 36]. Therefore, in a nodal-point semimetal, if one can tune its nodal point between type-I and type-II, not only the exotic transition of physical phenomena can be observed, but also the topological semimetal can serve as a device to control the variation of related physical quantities. However, although a lot of materials are found to be type-I *or* type-II nodal-point semimetals, no one can realize the transition between the two types. Moreover, no mechanism is proposed to realize the transition in real materials to date.

In this work, we propose that the transition can be realized in two-dimensional (2D) materials made of zigzag chains. Due to elasticity of the zigzag chains, these materials can sustain big external strains. Under the external strain, the orbital interactions between atoms drastically changes with structural deformation. Correspondingly, the slope signs of energy bands changes, and the topological phase transition between type-I and type-II Dirac/Weyl semimetals occurs. A tight-binding model based on a one-dimensional zigzag chain is used to explain the process of the transition. Two stable nanosheets, phosphorous nitride (PN) and arsenic nitride (AsN), are selected from a series of materials made of zigzag chains. Monolayer PN and AsN are intrinsic type-I and type-II Dirac semimetals, respectively, and they transit to another type of Dirac semimetals under strains. Finally, we propose a possible routine to synthesize the new 2D structures.

Figures 1(a) and 1(b) show type-I and type-II nodal points, respectively, and both are crossed by one blue and one red energy bands. The difference between them is the slope sign of the blue band. The nodal points transit from one type to another when the sign changes from negative to positive. We propose that the transition can be realized in a zigzag chain. Let us consider a zigzag chain locating on the *y-z* plane, as shown in Figs. 1(c-d). The primitive cell of the chain has two atoms, and its lattice constant is $a_0$. If each atom has one orbital[37], say $p_x$, one can use a tight-binding model to describe electronic property of the $p_x$ orbitals,

$$H_{p_x} = \sum_i \epsilon_{i,p_x} a_i^\dagger a_i + \sum_{i,j} t_{i,j,p_x} a_i^\dagger a_j , \qquad (1)$$

where $a_i^\dagger$ and $a_j$ represent the creation and the annihilation operators, respectively, $\epsilon_{i,p_x}$ is the site energy at site $i$, and $t_{i,j,p_x}$ is the hopping energy between sites $i$ and $j$. Here, we only consider the nearest-neighbor hopping $t_1$ and the next-nearest-neighbor hopping $t_2$, as shown in the left panels in Figs. 1(c-d). The two eigenvalues of Eq. (1) can be obtained by solving Schrodinger equation:

$$E_1 = \epsilon_{p_x} + t_2 \cos(k_y a) + \frac{1}{2} t_1 \cos(k_y a/2), \tag{2}$$

$$E_2 = \epsilon_{p_x} + t_2 \cos(k_y a) - \frac{1}{2} t_1 \cos(k_y a/2), \tag{3}$$

where $\epsilon_{p_x}$ is the site energy of $p_x$ orbital on each atom. When $t_1 > 0$, one can get $E_1 > E_2$ in the first Brillouin zone (BZ) $-\frac{\pi}{a} \leq k_y \leq \frac{\pi}{a}$. In this case, $E_1$ and $E_2$ represent antibonding and bonding states generated by $p_x$ orbitals, respectively. The slope signs of the energy bands corresponding to $E_1$ and $E_2$ are dependent on both $t_2$ and $k_y$. For example, if $t_2 > 0$, the slope sign of the $E_1$ band is always negative in the range $0 \leq k_y \leq \frac{\pi}{a}$, while that of the $E_2$ band close to $k_y = 0$ changes from positive to negative with the increase of $t_2$. The blue energy bands in Figs. 1(e-f) exhibit the effect of $t_2$ on the slope signs. It is noted that the value of $t_2$ is related to the distance of the next-nearest-neighbor atoms in the chain. The distance can be tuned by external strain, for example, a tensile/compressive strain along $y$ direction will increase/decrease the distance. This demonstrates that the slope signs of energy bands in a zigzag chain can be tuned by external strain applied on the chain, which can be verified in some one-dimensional (1D) zigzag chains.

If there is another orbital on each atom of the chain, say $p_z$, as shown in the right panels in Figs. 1(c-d), the orbital will also produce two energy bands, corresponding to an antibonding state $E_1' = \epsilon_{p_z} + t_2' \cos(k_y a) + \frac{1}{2} t_1' \cos(k_y a/2)$ and a bonding state $E_2' = \epsilon_{p_z} + t_2' \cos(k_y a) - \frac{1}{2} t_1' \cos(k_y a/2)$, respectively, where $\epsilon_{p_z}$ is the site energy of $p_z$ orbital, and $t_1'$ and $t_2'$ are the nearest-neighbor and the next-nearest-neighbor hoppings between $p_z$ orbitals, respectively. Similar to the case of $p_x$ orbital, the dispersions of the two bands are dependent on the parameters $t'_1$ and $t'_2$. Without loss of generality, one can assume that the variation of the slope signs of the two bands under the external strain is same to the case of $p_x$ orbital, i.e., the slope sign of $E'_1$ band does not change while that of $E'_2$ band varies with the strain. When $\epsilon_{p_z} < \epsilon_{p_x}$, the band $E'_1$ of $p_z$ orbital crosses with the $E_2$ band of $p_x$ orbital, and thus a nodal point is formed. When a tensile strain is applied on the chain along $y$ direction, the nodal point transits from type-I to type-II, as shown in Figs. 1(e-f), because of the variation of the slope sign of the $E_2$ band.

According to the mechanism of phase transition mentioned above, we propose a type of monolayer structures, made of 1D zigzag chains, to realize the phase transition between type-I and type-II nodal points in real 2D materials (in Fig. 2). The monolayer on $x$-$y$ plane in Fig. 2(b) is formed by linking zigzag chains in Fig. 2(a) along $y$ direction. The two neighboring chains locate up and down, and only the lower pink atoms in the upper chain link with the upper pink atoms in the lower chain. Thus, in the monolayer, the pink atoms $X$ are fourfold coordinate while the blue atoms $Y$ are twofold coordinate. The ratio of the two types of atoms is $X$:$Y$ = 1:1.

We performed first-principles calculations within the density functional theory (DFT) formalism as implemented in VASP[38]. The electron−electron interactions were treated within a generalized gradient approximation (GGA) in the form of Perdew−Burke−Ernzerhof (PBE) for the exchange−correlation functional[39, 40]. Electronic wave functions have been expanded using a plane-wave basis set with cut-off energy of 600 eV. The atomic positions were fully optimized by the conjugate gradient method[41], and the energy and force convergence criteria were set to be $10^{-6}$ eV and $10^{-3}$ eV/Å, respectively. Periodic boundary conditions were used with 20 Å vacuum layer in the direction perpendicular to the planes ($z$-direction), which ensures that the interaction between the periodic images of the sheet is negligible. The band structure was also calculated by using the Heyd-Scuseria-Ernzerhof 2006 (HSE06) approximation[42]. To account for the thermal stability, we carried out ab initio molecular dynamics (AIMD) simulations based on a canonical ensemble[43], for which a 4×4 supercell containing 72 atoms was used, and the AIMD simulations were performed with a Nose−Hoover thermostat from 300 to 1000 K.

We calculate band structures of a series of monolayers $XY$ ($X$, $Y$ = IV, V and VI elements) whose atomic structures are shown in Fig. 2, by using the first-principles methods. The PBE results indicate that only several monolayers are Dirac semimetals: CO, CS, PN, AsP are type-I Dirac semimetals, while AsN is a type-II Dirac semimetal (see Fig. S1). Then we access dynamical stabilities of these monolayer structures. Their phonon dispersions indicate that only monolayer CO, PN and AsN do not have soft modes in the phonon spectra (see Figs. S1), which means that only the three monolayers are dynamical stable. We further check band structures of monolayer CO, PN and AsN by using the HSE06 approximation, the results illustrate that monolayer CO is a semiconductor while monolayer PN and AsN are still type-I and type- II Dirac semimetals, respectively (see Fig. S2).

To make sure of thermal stability of monolayer PN and AsN, we perform AIMD simulations for the two structures in canonical ensemble. After heating up to the targeted temperature 800 and 600 K for 20 ps, respectively, we do not observe any structural decomposition in monolayer PN and AsN (see Figs. 3(c-f)). The reconstructions of the two structures occur at 1000 and 800 K during the 20 ps simulation, respectively. Therefore, monolayer PN and AsN have rather high thermal stabilities.

The primitive cells for monolayer PN and AsN are shown in a red dashed box in Fig. 2(c). After fully optimized, the lattice parameters for monolayer PN are $a_1$ = 3.79 Å and $b_1$ = 2.75 Å, respectively. The thickness of the structure is $d_1$ = 3.07 Å. The lengths of bonds P-N and P-P are 1.62 and 2.33 Å, respectively. The cohesive energy is 4.62 eV/atom, which can be comparable with the other PN monolayer (4.65 eV/atom)[44]. The lattice parameters for monolayer AsN are $a_2$ = 4.49 Å and $b_2$ = 2.90 Å, respectively. Its thickness is $d_2$ = 3.84 Å. The lengths of bonds As-N and As-As are 1.80 and 2.81 Å, respectively. The cohesive energy is 3.79 eV/atom. Both two monolayer structures belong to symmetry group PMMA (D2h-5).

The former calculations demonstrate that both monolayer PN and AsN could be good candidates for exploring phase transition between type-I and type-II Dirac semimetals[45-47]. We first calculate the intrinsic band structure of monolayer PN, as shown in Fig. 4(a). The conduction and valence bands cross on the Fermi level along the high symmetry line Γ-Y. The crossing is a type-I Dirac point because the slope signs of the two bands are inverse. A three-dimensional (3D) band structure around the Dirac point illustrates that it is a regular Dirac cone (see the inset in Fig. 4(a)). These indicate that monolayer PN is an intrinsic type-I Dirac semimetal. The velocities of Dirac electrons in the structure are approximate $5.0 \times 10^5 \ m/s$. Because the monolayer has both inversion and time-reversal symmetries[48], there are two symmetric Dirac points on $k_y$ axis in the first BZ, as shown in Fig. 4(d).

When a compressive strain is applied on monolayer PN along $z$ axis, the monolayer becomes thinner and the zigzag chains in the structure is stretched along x and y directions. The type-I Dirac points in monolayer PN gradually evolve into type-II Dirac points with the increase of the strains. Figure 4(b) presents the band structure of monolayer PN under a 2.0 GPa strain. Comparing Fig. 4(b) with Fig. 4(a), the points D1 and D3 in Fig. 4(a) shift up while the point D2 shifts down, which result in the slope of the red band through points D1 and D2 decreasing. The red band in Fig. 4(b) around the crossing point is nearly a flat band, and thus the crossing point is close to a critical Dirac point. When the compressive strain is increased to 2.1 GPa, the band structure is shown in Fig. 4(c). The points D1 continuously shifts up to D1' while the point D2 continuously shifts down to D2'. The slope of the band through points D1' and D2' becomes negative. Because the slopes of the two crossing bands have the same sign, the crossing point in Fig. 4(c) is a type-II Dirac point. The inset in Fig. 4(c) illustrates that type-II Dirac cone is strongly tilting, which results in the Dirac points locate at boundaries of the electron and hole pockets (see Fig. 4(e)). Figure S3 presents different topological edge states corresponding to type-I and type-II Dirac points.

To reveal origination of the phase transition in monolayer PN, we calculate its projected band structures, as shown in Fig. S4. The energy band through points D1 and D2 is attributed by $p_z$ orbitals of P and N atoms, while the band through point D3 is attributed by $p_x$ orbitals of P and N atoms. This is somewhat like the case in Figs. 1(c-d). Figure S5 presents the wavefunctions of the electron states at points D1, D3 and D1', D3' in Fig. 4. Under a compressive strain, the $p_z$ orbitals have a big change (see Figs. S6), although the zigzag chains are only slightly stretched. As a result, the slope of the corresponding band varies from a positive value to a negative one, and the phase transition occurs.

As mentioned above, monolayer AsN is an intrinsic type-II Dirac semimetals. Our calculations indicate that a compressive strain along $x$ and $y$ axes induces the structure transiting from type-II to type-I Dirac semimetal (see Fig. S6). It is noted that both monolayer PN and AsN have good elasticities because they are made of zigzag chains along both $x$ and $y$ directions. Thus, the structures can sustain a big tensile deformation. We calculate phonon dispersions of the structures under different strains, as shown in Fig. S6, and no soft mode is found in them. The variation of the lattice constants, bond lengths and thickness with the strain is given in Table S1. Because the As atom has a relatively larger spin-orbit coupling (SOC), we have also checked the effect of SOC on the Dirac points in the monolayer AsN. The result indicates that, in the case of SOC being considered, the crossing Dirac points are gapped with a very small gap ~ 1.3 meV (see Fig. S7).

The stability of monolayer PN discussed above has demonstrated the feasibility of synthesizing monolayer PN structure. However, which substrate could be a good choice? By comparing the lattice parameters between different material surfaces and monolayer PN, we find that the lattice parameters of Si (101) surface ($a'$ = 3.84Å, $b'$ = 5.43 Å) are close to those of a 1×2 supercell of monolayer PN ($a$ = 3.79Å, $b$ = 5.50 Å). The mismatch between them is less than 1.5%. It implies that the Si (101) surface could be a good substrate for growing the monolayer PN. The calculated adhesion energy between the monolayer PN and Si (101) substrate as shown in Fig. 5 is -0.08 eV/atom, which can be compare with some 2D materials on their substrates [49, 50]. We do believe the newly monolayer PN can be synthesized on the Si substrate. Moreover, the effect of the substrate on the band structure of monolayer PN is weak (see Fig. S8), and thus one can use the interesting electron properties of monolayer PN directly on the substrate.

In summary, we propose a mechanism to realize the transition between type-I and type-II Dirac semimetals by a tight-binding model. The mechanism is confirmed in some monolayers, such as monolayer PN and AsN, made of zigzag chains. Monolayer PN is an intrinsic type-I Dirac semimetal while monolayer AsN is a type-II Dirac semimetals. Due to elasticity of the zigzag chains, these materials can sustain big external strains. Monolayer PN transits to a type-II Dirac semimetal when a compressive strain is applied on the 2D material, because the orbital interactions between atoms drastically changes with structural deformation. The monolayer structure could be synthesized on a Si (101) surface. On the other hand, the monolayer AsN transits from a type-II to a type-I semimetal under strain.

Our work proposes not only a series of monolayer structures made of zigzag chains but also a mechanism to realize transitions between type-I and type-II semimetals. Because the transitions induce corresponding significant variations of electronic, optical and magnetic properties, the monolayers provide a platform to observe the important evolution processes as well as to their relative applications.

This work was supported by the National Natural Science Foundation of China (No.11874314).

The data that supports the findings of this study are available within the article and its supplementary material.

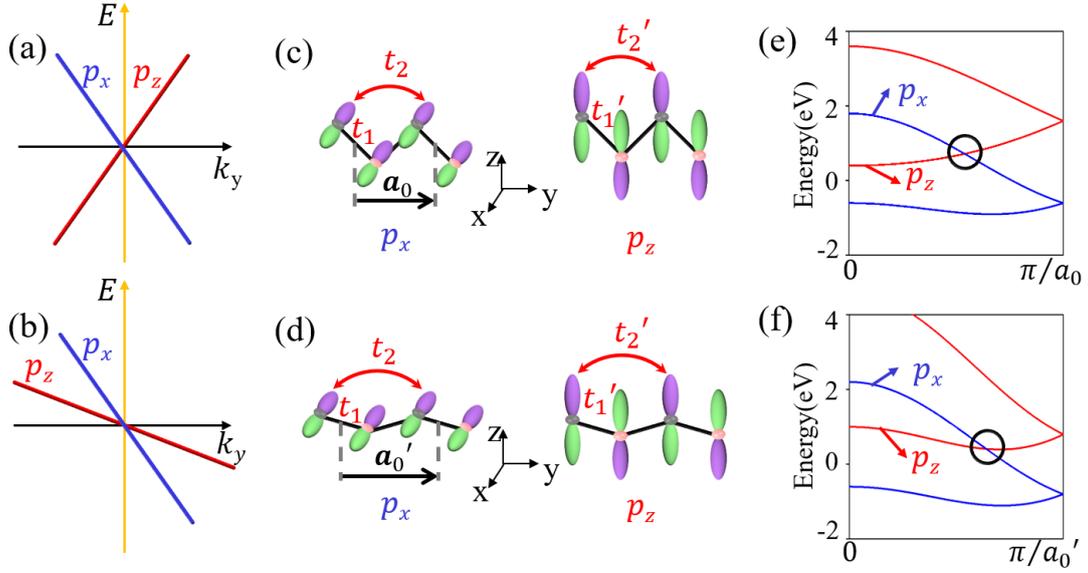

Figure 1. Schematic views of (a) type-I and (b) type-II Dirac points, which are crossings of a blue band and a red band. (c) Two possible orbitals in a zigzag chain locating on the $y$-$z$ plane: $p_x$ orbital (left panel) and $p_z$ orbital (right panel). $a_0$ is the lattice constant of the structure. (d) The same to (c) when the zigzag chain is elongated along $y$ direction. The lattice constant is changed to $a_0'$. In (c-d), $t_1$ and $t_1'$ represent the nearest-neighbor hopping energies of $p_x$ and $p_z$ orbitals, respectively; $t_2$ and $t_2'$ represent the next-nearest-neighbor hopping energies of $p_x$ and $p_z$ orbitals, respectively. (e-f) Band structures of the zigzag chains in (c-d) based on Eq. (1), respectively. The parameters for (e) are $t_1= 0.6$ eV, $t_1'= 0.8$ eV, $t_2= 0.3$ eV, $t_2'= 0.1$ eV, $\epsilon_{pz} = 1.8$ eV, $\epsilon_{px}= 0$ eV; while the parameters for (f) are $t_1= 0.7$ eV, $t_1'= 0.9$ eV, $t_2= 0.4$ eV, $t_2'= 0.5$ eV, $\epsilon_{pz} = 1.8$ eV, $\epsilon_{px} = 0$ eV。One can note that the crossing in the circle in (e) is a type-I Dirac point, while that in (f) is a type-II Dirac point.

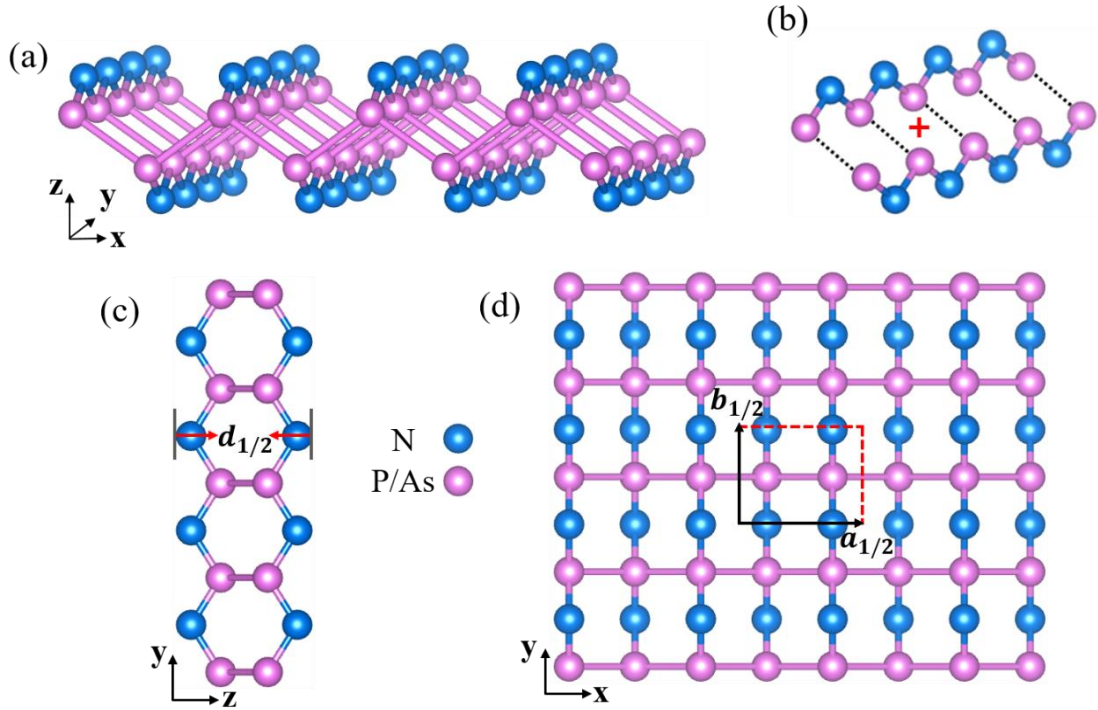

Figure 2. (a) Atomic structures of monolayer PN and AsN, where the blue atoms are N atoms while the purple atoms are P or As atoms. This structure can be viewed as PN (or AsN) zigzag chains along *y* direction link together along *x* direction, as shown in (b). The P (or As) atoms link together and then form zigzag chains along *x* direction. (c) Side view of the monolayer structure, where $d_{1/2}$ represent the thickness of the monolayer PN and AsN, respectively. (d) Top view of the monolayer structure. The primitive cell is shown in the red dashed box in (c). $a_{1/2}$ and $b_{1/2}$ represent the lattice constants of PN and AsN, respectively.

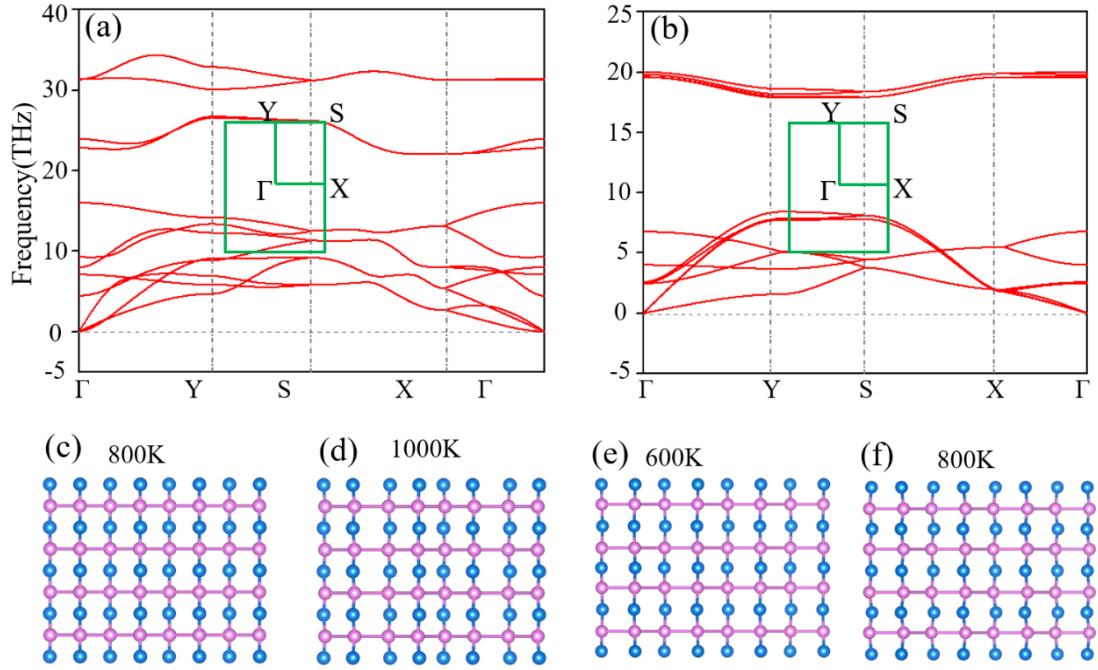

Figure 3. (a) Phonon dispersion of monolayer PN. The first BZ is given in the inset. Snapshots of the equilibrium structures of monolayer PN at temperatures of (c) 800 and (d) 1000 K after 20 ps AIMD simulations. (b) Phonon dispersion of monolayer AsN. The first BZ is given in the inset. Snapshots of the equilibrium structures of monolayer AsN at temperatures of (e) 600 and (f) 800 K after 20 ps AIMD simulations.

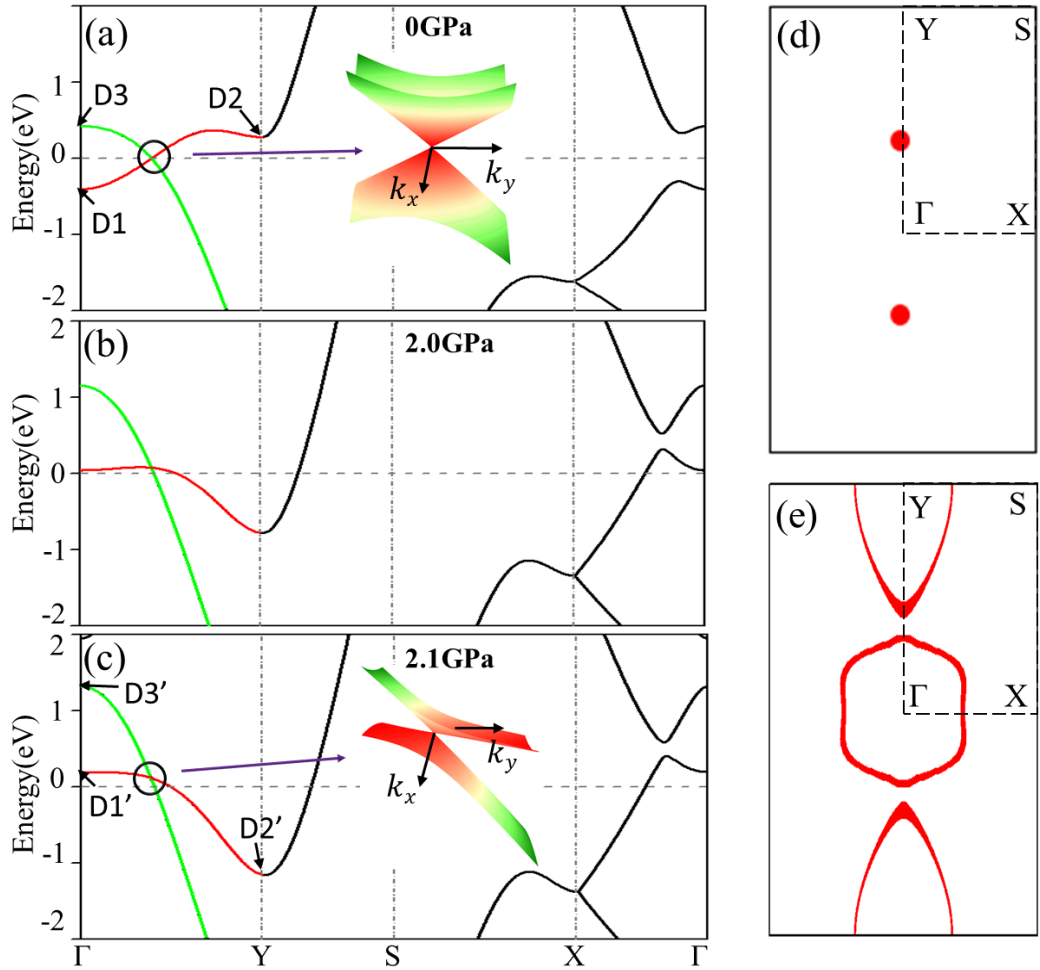

Figure 4. Band structures of monolayer PN when compressive strains (a) 0 GPa, (b) 2.0 GPa and (c) 2.1GPa are applied on the structure along *z* direction. The insets in (a) and (c): 3D band structures around the type-I and type-II Dirac points in (a) and (c), respectively. (d-e) The Fermi surfaces corresponding to the band structures in (a) and (c), respectively.

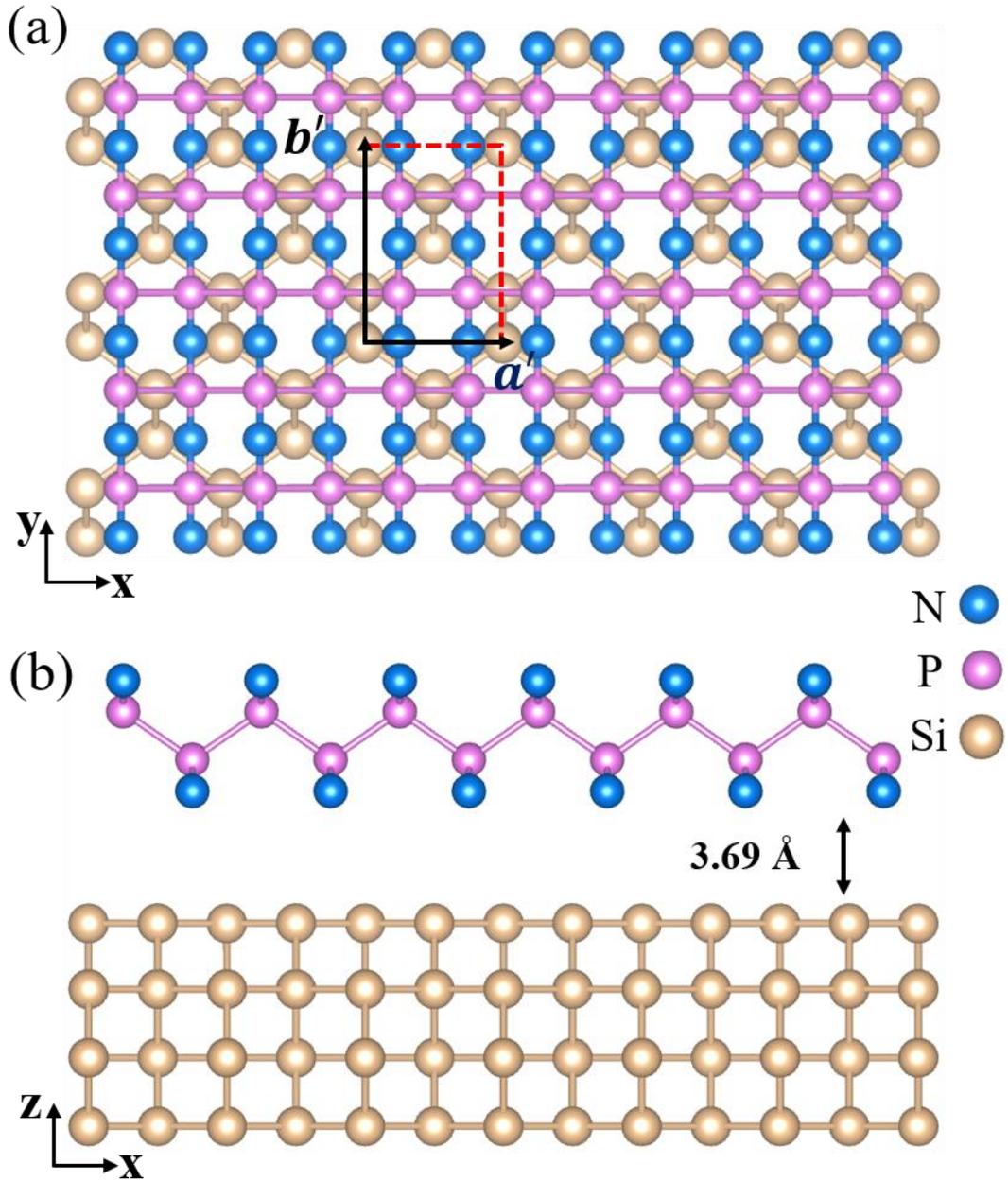

Figure 5. Top view (a) and side view (b) of monolayer PN on a Si (101) substrate. The primitive cell is shown in the dashed box in (a), where $a'$ and $b'$ represent the lattice constants of the monolayer and its substrate. The distance between the Si (101) substrate and the PN monolayer is 3.69 Å.